\@twosidetrue
\@twocolumntrue
\@mparswitchtrue
\def\ds@draft{\overfullrule 5pt}
\def\ds@twocolumn{\@twocolumntrue}
\def\ds@onecolumn{\@twocolumnfalse}
\newif\ifSFB@landscape
\def\ds@landscape{\SFB@landscapetrue}
\newif\ifSFB@galley
\def\ds@galley{\SFB@galleytrue}
\newif\ifSFB@referee
\def\ds@referee{%
 \SFB@refereetrue
 \@twocolumnfalse
}
\@options
\ifSFB@referee
 
\else
 
\fi
\if@twocolumn
  \def\@normalsize{\@setsize\normalsize{11pt}\ixpt\@ixpt
   \abovedisplayskip 6pt plus 2pt minus 2pt
   \belowdisplayskip \abovedisplayskip
   \abovedisplayshortskip 6pt plus 2pt
   \belowdisplayshortskip \abovedisplayshortskip
   \let\@listi\@listI}
 \else
 \ifSFB@referee
  \def\@normalsize{\@setsize\normalsize{14pt}\xiipt\@xiipt
   \abovedisplayskip 4pt plus 1pt minus 1pt
   \belowdisplayskip \abovedisplayskip
   \abovedisplayshortskip 4pt plus 1pt
   \belowdisplayshortskip \abovedisplayshortskip
   \let\@listi\@listI}
 \else
  \def\@normalsize{\@setsize\normalsize{12pt}\ixpt\@ixpt
   \abovedisplayskip 4pt plus 1pt minus 1pt
   \belowdisplayskip \abovedisplayskip
   \abovedisplayshortskip 4pt plus 1pt
   \belowdisplayshortskip \abovedisplayshortskip
   \let\@listi\@listI}
 \fi
\fi
\def\small{\@setsize\small{10pt}\viiipt\@viiipt
 \abovedisplayskip 4pt plus 1pt minus 1pt
 \belowdisplayskip \abovedisplayskip
 \abovedisplayshortskip 4pt plus 1pt
 \belowdisplayshortskip \abovedisplayshortskip
 \def\@listi{\leftmargin\leftmargini
  \topsep 2pt plus 1pt minus 1pt
  \parsep \z@
  \itemsep 2pt}}
\def\footnotesize{\@setsize\footnotesize{10pt}\viiipt\@viiipt
 \abovedisplayskip 4pt plus 1pt minus 1pt
 \belowdisplayskip \abovedisplayskip
 \abovedisplayshortskip 4pt plus 1pt
 \belowdisplayshortskip \abovedisplayshortskip
 \def\@listi{\leftmargin\leftmargini
  \topsep 2pt plus 1pt minus 1pt
  \parsep \z@
  \itemsep 2pt}}
\def\scriptsize{\@setsize\scriptsize{8pt}\viipt\@viipt}
\def\tiny{\@setsize\tiny{6pt}\vpt\@vpt}
\if@twocolumn
  \def\large{\@setsize\large{11pt}\xpt\@xpt}
 \else
  \def\large{\@setsize\large{12pt}\xpt\@xpt}
 \fi
\def\Large{\@setsize\Large{14pt}\xiipt\@xiipt}
\def\LARGE{\@setsize\LARGE{17pt}\xivpt\@xivpt}
\def\huge{\@setsize\huge{20pt}\xviipt\@xviipt}
\def\Huge{\@setsize\huge{25pt}\xxpt\@xxpt}
\normalsize

\if@twocolumn
\else
 \ifSFB@referee
  \oddsidemargin \z@  \evensidemargin \z@
 \else
 \fi
\fi
 \topmargin \z@
\newdimen\SFB@measure
\textwidth \SFB@measure
\ifSFB@landscape
 \textwidth \textheight
 \textheight \SFB@measure
\fi
\ifSFB@referee
\fi
\skip\footins 19.5pt plus 12pt minus 1pt

\parskip \z@ plus .1pt
\@beginparpenalty -\@lowpenalty
\@endparpenalty -\@lowpenalty
\@itempenalty -\@lowpenalty
\newcounter{part}
\newcounter {section}
\newcounter {subsection}[section]
\newcounter {subsubsection}[subsection]
\newcounter {paragraph}[subsubsection]
\newcounter {subparagraph}[paragraph]
\def\thepart          {\arabic{part}}
\def\thesection       {\arabic{section}}

\def\part{\par \addvspace{4ex}\@afterindentfalse
 \secdef\@part\@spart}
\def\@part[#1]#2{\ifnum \c@secnumdepth >\m@ne
  \refstepcounter{part}
  \addcontentsline{toc}{part}{Part \thepart: #1}
 \else \addcontentsline{toc}{part}{#1}
 \fi
 {\parindent 0pt \raggedright
  \ifnum \c@secnumdepth >\m@ne
   \large\rm PART
   \ifcase\thepart \or ONE \or TWO \or THREE \or FOUR \or FIVE
    \or SIX \or SEVEN \or EIGHT \or NINE \or TEN \else \fi
   \par \nobreak
  \fi
  \LARGE \rm #2 \markboth{}{}\par }
 \nobreak \vskip 3ex \@afterheading}
\def\@spart#1{{\parindent 0pt \raggedright
  \LARGE \rm #1\par}
 \nobreak \vskip 3ex \@afterheading}

\def\section{\@startsection {section}{1}{\z@}
 {-24pt plus -12pt minus -1pt}
 {6pt}
 {\SFB@hangraggedright\normalsize\bf}}
\def\subsection{\@startsection{subsection}{2}{\z@}
 {-18pt plus -9pt minus -1pt}
 {6pt}
 {\SFB@hangraggedright\large\bf}}
\def\subsubsection{\@startsection{subsubsection}{3}{\z@}
 {-18pt plus -9pt minus -1pt}
 {6pt}
 {\SFB@hangraggedright\normalsize\it}}
\def\paragraph{\@startsection{paragraph}{4}{\z@}
 {12pt plus 2.25pt minus 1pt}{-0.5em}{\normalsize\bf}}
\def\subparagraph{\@startsection{subparagraph}{5}{\parindent}
 {12pt plus 2.25pt minus 1pt}{-0.5em}{\normalsize\it}}
\def\SFB@hangraggedright{\rightskip\@flushglue \let\\=\newline}
\def\@sect#1#2#3#4#5#6[#7]#8{%
 \ifnum #2>\c@secnumdepth
  \def\@svsec{}%
 \else
  \refstepcounter{#1}
  \ifnum #2=\@ne
   \ifSFB@appendix \edef\@svsec{}%
             \else \edef\@svsec{\csname the#1\endcsname\hskip 1em}%
   \fi
  \else \edef\@svsec{\csname the#1\endcsname\hskip 1em}%
  \fi
 \fi
 \@tempskipa #5\relax
 \ifdim \@tempskipa>\z@
  \begingroup #6\relax
   \ifnum #2=\@ne
    \ifSFB@appendix
     \@hangfrom{\hskip #3\relax\@svsec}{\interlinepenalty \@M
      APPENDIX \csname the#1\endcsname:\hskip 0.5em\uppercase{#8}\par}%
    \else
     \@hangfrom{\hskip #3\relax\@svsec}{\interlinepenalty \@M
      \uppercase{#8}\par}%
    \fi
   \else
    \@hangfrom{\hskip #3\relax\@svsec}{\interlinepenalty \@M #8\par}%
   \fi
  \endgroup
  \csname #1mark\endcsname{#7}%
  \addcontentsline{toc}{#1}{\ifnum #2>\c@secnumdepth \else
   \protect\numberline{\csname the#1\endcsname}\fi #7}%
 \else
  \def\@svsechd{#6\hskip #3\@svsec \ifnum #2=\@ne\uppercase{#8}\else #8\fi
  \csname #1mark\endcsname{#7}
  \addcontentsline{toc}{#1}{\ifnum #2>\c@secnumdepth \else
   \protect\numberline{\csname the#1\endcsname}\fi#7}}%
 \fi
 \@xsect{#5}}

\newif\ifSFB@appendix
\def\appendix{\par
 \SFB@appendixtrue
 \setcounter{section}{0}
 \def\thesection{A\arabic{section}}
 \setcounter{equation}{0}
 \def\theequation{A\arabic{equation}}
 \setcounter{figure}{0}
 \def\thefigure{A\@arabic\c@figure}
 \setcounter{table}{0}
 \def\thetable{A\@arabic\c@table}
}

\newskip\@indentskip
\newskip\smallindent
\newskip\@footindent
\newskip\@leftskip
\@footindent=\smallindent
\@leftskip=\z@

\leftmargini   \@indentskip
\leftmargin\leftmargini
\labelwidth\leftmargini\advance\labelwidth-\labelsep
\def\makeRRlabel#1{\hss\llap{#1}}
\def\@listI{\leftmargin\leftmargini
 \parsep \z@
 \topsep 6pt plus 1pt minus 1pt
 \itemsep \z@ plus .1pt
}
\let\@listi\@listI
\@listi
\def\@listii{\leftmargin\leftmarginii
 \labelwidth\leftmarginii\advance\labelwidth-\labelsep
 \topsep 6pt plus 1pt minus 1pt
 \parsep \z@
 \itemsep \z@ plus .1pt
}
\def\@listiii{\leftmargin\leftmarginiii
 \labelwidth\leftmarginiii\advance\labelwidth-\labelsep
 \topsep 6pt plus 1pt minus 1pt
 \parsep \z@
 \partopsep \z@
 \itemsep \topsep
}
\def\@listiv{\leftmargin\leftmarginiv
 \labelwidth\leftmarginiv\advance\labelwidth-\labelsep
}
\def\@listv{\leftmargin\leftmarginv
 \labelwidth\leftmarginv\advance\labelwidth-\labelsep
}
\def\@listvi{\leftmargin\leftmarginvi
 \labelwidth\leftmarginvi\advance\labelwidth-\labelsep
}
\def\itemize{\ifnum \@itemdepth >3 \@toodeep
  \else \advance\@itemdepth \@ne
   \edef\@itemitem{labelitem\romannumeral\the\@itemdepth}%
   \list{\csname\@itemitem\endcsname}%
    {\let\makelabel\makeRRlabel}%
  \fi}

\def\enumerate{\ifnum \@enumdepth >3 \@toodeep \else
  \advance\@enumdepth \@ne
  \edef\@enumctr{enum\romannumeral\the\@enumdepth}%
 \fi
 \@ifnextchar [{\@enumeratetwo}{\@enumerateone}%
}
\def\@enumeratetwo[#1]{%
 \list{\csname label\@enumctr\endcsname}%
  {\settowidth\labelwidth{[#1]}
   \leftmargin\labelwidth \advance\leftmargin\labelsep
   \usecounter{\@enumctr}
   \let\makelabel\makeRRlabel}
}
\def\@enumerateone{%
 \list{\csname label\@enumctr\endcsname}%
  {\usecounter{\@enumctr}
   \let\makelabel\makeRRlabel}}

\def\theenumi{(\roman{enumi})}

\def\theenumii{(\alph{enumii})}
\def\p@enumii{\theenumi}

\def\theenumiii{(\arabic{enumiii})}
\def\p@enumiii{\theenumi(\theenumii)}

\def\p@enumiv{\p@enumiii\theenumiii}
\def\description{\list{}{\labelwidth\z@ \itemindent-\leftmargin
  \leftmargin 1em
  \itemindent-1em
}}

\def\verse{\let\\=\@centercr
 \list{}{\itemsep\z@
  \itemindent -\@indentskip
  \listparindent \itemindent
  \rightmargin\leftmargin
  \advance\leftmargin \@indentskip}\item[]}

\def\quotation{\list{}{\listparindent \smallindent
  \leftmargin\z@\rightmargin\leftmargin
  \parsep 0pt plus 1pt}\item[]\small}

\def\quote{\list{}{\leftmargin\z@\rightmargin\leftmargin}\item[]\small}

\def\@begintheorem#1#2{\rm \trivlist \item[\hskip \labelsep{\bf #1\ #2.}]}
\def\@opargbegintheorem#1#2#3{\rm \trivlist
  \item[\hskip \labelsep{\bf #1\ #2.\ (#3)}]}
\def\@endtheorem{\endtrivlist}
\@namedef{proof*}{\rm \trivlist \item[\hskip \labelsep{\it Proof.}]}
\@namedef{endproof*}{\endtrivlist}

\def\titlepage{\@restonecolfalse\if@twocolumn\@restonecoltrue\onecolumn
  \else \newpage \fi \thispagestyle{empty}\c@page\z@}
\def\endtitlepage{\if@restonecol\twocolumn \else \newpage \fi}

\def\tabular{\def\@halignto{}
 \def\hline{\noalign{\ifnum0=`}\fi
  \vskip 3pt
  \hrule \@height \arrayrulewidth
  \vskip 3pt
  \futurelet \@tempa\@xhline}
 \def\fullhline{\noalign{\ifnum0=`}\fi
  \vskip 3pt
  \hrule \@height \arrayrulewidth
  \vskip 3pt
  \futurelet \@tempa\@xhline}
 \def\@xhline{\ifx\@tempa\hline
   \vskip -6pt
   \vskip \doublerulesep
  \fi
  \ifnum0=`{\fi}}
  \def\@arrayrule{\@addtopreamble{\hskip -.5\arrayrulewidth
                                  \hskip .5\arrayrulewidth}}
\@tabular
}
\tabbingsep \labelsep

\skip\@mpfootins = \skip\footins

\fboxrule = \arrayrulewidth

\def\maketitle{\par
 \begingroup
  \if@twocolumn
   \twocolumn[\vspace*{17pt}\@maketitle]
  \else
   \newpage
   \global\@topnum\z@
   \@maketitle
  \fi
  \thispagestyle{titlepage}
 \endgroup
 \let\maketitle\relax
 \let\@maketitle\relax
 \gdef\@author{}
 \gdef\@title{}
 \let\thanks\relax
}
\def\and{\end{author@tabular}\vskip 6pt\par
 \begin{author@tabular}[t]{@{}l@{}}}
\def\@maketitle{\newpage
 \vspace*{7pt}
 {\raggedright \sloppy
  {\huge \bf \@title \par}
  \vskip 23pt
  {\LARGE
   \begin{author@tabular}[t]{@{}l@{}}\@author
   \end{author@tabular}\par}
  \vskip 22pt
 }
 \par\noindent
 {\small \@date \par}
 \vskip 22pt
}
\def\abstract{\if@twocolumn
  \start@SFBbox\@abstract
 \else
  \@abstract
 \fi}
\def\endabstract{\if@twocolumn
   \endlist\finish@SFBbox
 \else
  \endlist
 \fi}
\def\@abstract{\list{}{\leftmargin 10.5pc\rightmargin\z@
  \parsep 0pt plus 1pt}\item[]\normalsize{\bf ABSTRACT}\\\large} 
\newif\ifSFB@keywords
\def\keywords{\if@twocolumn
  \start@SFBbox\@keywords
 \else
  \@keywords
 \fi
}
\def\@keywords{\list{}{\leftmargin 10.5pc\rightmargin\z@
  \parsep 0pt plus 1pt}\item[]\large{\bf Key words: }}
\def\endkeywords{\if@twocolumn
  \endlist\addvspace{37pt}\finish@SFBbox
 \else
  \endlist
 \fi
 \@thanks
 \gdef\@thanks{}
 \SFB@keywordstrue
}
\def\nokeywords{\ifSFB@keywords\else
 \if@twocolumn \start@SFBbox\addvspace{37pt}\finish@SFBbox \fi
 \@thanks
 \gdef\@thanks{}\fi
}

\def\author@tabular{\def\@halignto{}\@authortable}
\let\endauthor@tabular=\endtabular
\def\author@tabcrone{{\ifnum0=`}\fi\@xtabularcr[-7pt]\small\it
 \let\\=\author@tabcrtwo\ignorespaces}
\def\author@tabcrtwo{{\ifnum0=`}\fi\@xtabularcr[-7pt]\small\it
 \let\\=\author@tabcrtwo\ignorespaces}
\def\@authortable{\leavevmode \hbox \bgroup $\let\@acol\@tabacol
 \let\@classz\@tabclassz \let\@classiv\@tabclassiv
 \let\\=\author@tabcrone \ignorespaces \@tabarray}

\def\start@SFBbox{\@next\@currbox\@freelist{}{}%
 \global\setbox\@currbox
 \vbox\bgroup
  \hsize \textwidth
  \@parboxrestore
}
\def\finish@SFBbox{\par\vskip -\dbltextfloatsep
  \egroup
  \global\count\@currbox\tw@
  \global\@dbltopnum\@ne
  \global\@dbltoproom\maxdimen\@addtodblcol
  \global\vsize\@colht
  \global\@colroom\@colht
}

\mark{{}{}}
\gdef\@author{\mbox{}}
\def\author{\@ifnextchar [{\@authortwo}{\@authorone}}
\def\@authortwo[#1]#2{\gdef\@author{#2}\gdef\@shortauthor{#1}}
\def\@authorone#1{\gdef\@author{#1}\gdef\@shortauthor{#1}}
\gdef\@shortauthor{}
\gdef\@title{\mbox{}}
\def\title{\@ifnextchar [{\@titletwo}{\@titleone}}
\def\@titletwo[#1]#2{\gdef\@title{#2}\gdef\@shorttitle{#1}}
\def\@titleone#1{\gdef\@title{#1}\gdef\@shorttitle{#1}}
\gdef\@shorttitle{}
\def\volume#1{\gdef\@volume{#1}}
\gdef\@volume{000}
\def\microfiche#1{\gdef\@microfiche{#1}}
\gdef\@microfiche{}
\def\pagerange#1{\gdef\@pagerange{#1}}
\gdef\@pagerange{000--000}
\def\journal#1{\gdef\@journal{#1}}
\gdef\@journal{{Mon.\ Not.\ R.\ Astron.\ Soc.} {\bf \@volume}, \@pagerange\
  (\number\year) \@microfiche}
\def\ps@headings{\let\@mkboth\markboth
 \def\@oddhead{\Large \hfill \it \@shorttitle \hspace{1.5em}\rm \thepage}
 \def\@oddfoot{}
 \def\@evenhead{\Large \thepage \hspace{1.5em}\it \@shortauthor \hfill}
 \def\@evenfoot{}
 \def\sectionmark##1{\markboth{##1}{}}
 \def\subsectionmark##1{\markright{##1}}}
\def\ps@myheadings{\let\@mkboth\@gobbletwo
 \def\@oddhead{\Large \it \rightmark \hfill \rm \thepage}
 \def\@oddfoot{}
 \def\@evenhead{\Large \it \leftmark \hfill \rm \thepage}
 \def\@evenfoot{}
 \def\sectionmark##1{}
 \def\subsectionmark##1{}}
\def\ps@titlepage{\let\@mkboth\@gobbletwo
 \def\@oddhead{\footnotesize\@journal\hfill}
 \def\@oddfoot{}
 \def\@evenhead{\footnotesize\@journal\hfill}
 \def\@evenfoot{}
 \def\sectionmark##1{}
 \def\subsectionmark##1{}}

\def\@pnumwidth{1.55em}
\def\@tocrmarg {2.55em}
\def\@dotsep{4.5}
\def\@undottedtocline#1#2#3#4#5{\ifnum #1>\c@tocdepth
 \else
  \vskip \z@ plus .2pt
  {\hangindent #2\relax
   \rightskip \@tocrmarg \parfillskip -\rightskip
   \parindent #2\relax \@afterindenttrue
   \interlinepenalty\@M \leavevmode
   \@tempdima #3\relax #4\nobreak \hfill \nobreak
   \hbox to\@pnumwidth{\hfil\rm \ }\par}\fi}
\def\tableofcontents{\@restonecolfalse
 \if@twocolumn\@restonecoltrue\onecolumn\fi
 \section*{CONTENTS} \@starttoc{toc}
 \if@restonecol\twocolumn\fi \par\vspace{12pt}}
\def\l@part#1#2{\addpenalty{-\@highpenalty}
 \addvspace{2.25em plus 1pt}
 \begingroup
  \parindent \z@ \rightskip \@pnumwidth
  \parfillskip -\@pnumwidth
  {\normalsize\rm
   \leavevmode \hspace*{3pc}
   #1\hfil \hbox to\@pnumwidth{\hss \ }}\par
   \nobreak \global\@nobreaktrue
   \everypar{\global\@nobreakfalse\everypar{}}\endgroup}
\def\l@section#1#2{\addpenalty{\@secpenalty}
 \@tempdima 1.5em
 \begingroup
  \parindent \z@ \rightskip \@pnumwidth
  \parfillskip -\@pnumwidth \rm \leavevmode
  \advance\leftskip\@tempdima \hskip -\leftskip
  #1\nobreak\hfil \nobreak\hbox to\@pnumwidth{\hss \ }\par
 \endgroup}
\def\l@subsection{\@undottedtocline{2}{1.5em}{2.3em}}
\def\l@subsubsection{\@undottedtocline{3}{3.8em}{3.2em}}
\def\l@paragraph{\@undottedtocline{4}{7.0em}{4.1em}}
\def\l@subparagraph{\@undottedtocline{5}{10em}{5em}}
\def\listoffigures{\@restonecolfalse
 \if@twocolumn\@restonecoltrue\onecolumn\fi
 \section*{LIST OF FIGURES\@mkboth{LIST OF FIGURES}{LIST OF FIGURES}}
 \@starttoc{lof} \if@restonecol\twocolumn\fi}
\def\l@figure{\@undottedtocline{1}{1.5em}{2.3em}}
\def\listoftables{\@restonecolfalse
 \if@twocolumn\@restonecoltrue\onecolumn\fi
 \section*{LIST OF TABLES\@mkboth{LIST OF TABLES}{LIST OF TABLES}}
 \@starttoc{lot} \if@restonecol\twocolumn\fi}
\let\l@table\l@figure

\def\thebibliography#1{\section*{REFERENCES}
 \addcontentsline{toc}{section}{REFERENCES}
 \list{}{\labelwidth\z@
         \leftmargin 1.5em
	 \itemsep \z@
	 \itemindent-\leftmargin}
 \small\raggedright
 \parindent\z@
 \parskip\z@ plus .1pt\relax
 \def\newblock{\hskip .11em plus .33em minus .07em}
 \sloppy\clubpenalty4000\widowpenalty4000
 \sfcode`\.=1000\relax
}

\def\@biblabel#1{\hspace*{\labelsep}[#1]}

\newif\if@restonecol
\def\theindex{\section*{INDEX}
 \addcontentsline{toc}{section}{INDEX}
 \footnotesize \parindent\z@ \parskip\z@ plus .1pt\relax
 \let\item\@idxitem}
\def\@idxitem{\par\hangindent 1em}

\def\endtheindex{\if@restonecol\onecolumn\else\clearpage\fi}

\def\footnoterule{\kern-3\p@ \hrule width 12pc height \z@ \kern 3\p@}

\def\@fnsymbol#1{\ifcase#1\or \mbox{$\star$}\or \dagger\or \ddagger\or
   \S \or \P \or \|\or **\or \dagger\dagger
   \or \ddagger\ddagger\or \S\S\or \P\P\or \|\|\else ***
   \fi\relax}

\long\def\@makefntext#1{\parindent 1em\noindent
  $^{\@thefnmark}$\hspace{4pt}#1}

\newcounter{table}
\def\thetable{\@arabic\c@table}
\def\fps@table{tbp}
\def\ftype@table{1}
\def\ext@table{lot}
\def\fnum@table{Table \thetable}
\def\table{\let\@makecaption=\SFB@maketablecaption\@float{table}}
\let\endtable\end@float
\@namedef{table*}{\let\@makecaption=\SFB@maketablecaption\@dblfloat{table}}
\@namedef{endtable*}{\end@dblfloat}

\newcounter{figure}
\def\thefigure{\@arabic\c@figure}
\def\fps@figure{tbp}
\def\ftype@figure{2}
\def\ext@figure{lof}
\def\fnum@figure{Figure \thefigure}
\def\figure{\let\@makecaption=\SFB@makefigurecaption\@float{figure}}
\let\endfigure\end@float
\@namedef{figure*}{\let\@makecaption=\SFB@makefigurecaption\@dblfloat{figure}}
\@namedef{endfigure*}{\end@dblfloat}

\long\def\SFB@makefigurecaption#1#2{\vskip 6pt
 \setbox\@tempboxa\hbox{\small{\bf #1.} #2}
 \ifdim \wd\@tempboxa >\hsize
  \small{\bf #1.} #2\par
 \else
  \hbox to\hsize{\hfil\box\@tempboxa\hfil}
 \fi
 \vskip 6pt
}
\long\def\SFB@maketablecaption#1#2{\vskip 6pt
 \setbox\@tempboxa\hbox{\small{\bf #1.} #2}
 \ifdim \wd\@tempboxa >\hsize
  \small{\bf #1.} #2\par
 \else
  \hbox to\hsize{\box\@tempboxa\hfill}
 \fi
 \vskip 6pt
}

\def\caption{\@ifstar{\SFB@caption\@captype}%
 {\refstepcounter\@captype \@dblarg{\@caption\@captype}}%
}
\long\def\SFB@caption#1#2{
 \begingroup
  \@parboxrestore
  \normalsize
  \@makecaption{\csname fnum@#1\endcsname}{\ignorespaces #2}\par
 \endgroup}

\def\@cite#1#2{(#1\if@tempswa , #2\fi)}
\def\@biblabel#1{}

\newlength{\bibhang}
\def\@citex[#1]#2{\if@filesw\immediate\write\@auxout{\string\citation{#2}}\fi
  \def\@citea{}\@cite{\@for\@citeb:=#2\do
    {\@citea\def\@citea{; }\@ifundefined
       {b@\@citeb}{{\bf ?}\@warning
       {Citation `\@citeb' on page \thepage \space undefined}}%
{\csname b@\@citeb\endcsname}}}{#1}}
\let\@internalcite\cite
\def\cite{\def\citename##1{##1}\@internalcite}
\def\shortcite{\def\citename##1{}\@internalcite}

\def\[{\relax\ifmmode\@badmath\else\begin{trivlist}\item[]\leavevmode
  \hbox to\linewidth\bgroup$
  \displaystyle
  \hskip\mathindent\bgroup\fi}

\def\]{\relax\ifmmode \egroup $\hfil
       \egroup \end{trivlist}\else \@badmath \fi}

\def\equation{\refstepcounter{equation}\trivlist \item[]\leavevmode
  \hbox to\linewidth\bgroup $
  \displaystyle
\hskip\mathindent}

\def\endequation{$\hfil
           \displaywidth\linewidth\@eqnnum\egroup \endtrivlist}

\def\eqnarray{\stepcounter{equation}\let\@currentlabel=\theequation
\global\@eqnswtrue
\global\@eqcnt\z@\tabskip\mathindent\let\\=\@eqncr
\abovedisplayskip\topsep\ifvmode\advance\abovedisplayskip\partopsep\fi
\belowdisplayskip\abovedisplayskip
\belowdisplayshortskip\abovedisplayskip
\abovedisplayshortskip\abovedisplayskip
$$\halign
to \linewidth\bgroup\@eqnsel\hskip\@centering$\displaystyle\tabskip\z@
  {##}$&\global\@eqcnt\@ne \hskip 2\arraycolsep \hfil${##}$\hfil
  &\global\@eqcnt\tw@ \hskip 2\arraycolsep $\displaystyle{##}$\hfil
   \tabskip\@centering&\llap{##}\tabskip\z@\cr}

\def\endeqnarray{\@@eqncr\egroup
 \global\advance\c@equation\m@ne$$\global\@ignoretrue}

\newdimen\mathindent
\mathindent = \z@

\def\today{\number\day\ \ifcase\month\or
  January\or February\or March\or April\or May\or June\or
  July\or August\or September\or October\or November\or December
 \fi \ \number\year}

\ps@headings
\ifSFB@galley
 \ps@empty
\fi
\ifSFB@referee
\fi
\if@twocolumn
 \twocolumn
\else
 \onecolumn
\fi
\documentstyle[onecolumn]{mn}

\title{Weighted Bias and Galaxy Clustering}
\author[Paolo Catelan, Peter Coles,
Sabino Matarrese and Lauro Moscardini]{Paolo Catelan$^{1}$,
Peter Coles$^{2}$,
Sabino Matarrese$^{3}$ and Lauro Moscardini$^{4}$ \\
$^{1}$ Scuola Internazionale Superiore di Studi Avanzati, SISSA,
via Beirut 2--4, I--34013 Trieste, Italy\\
$^{2}$ Astronomy Unit, School of Mathematical Sciences,
Queen Mary and Westfield College, Mile End Road, London E1 4NS\\
$^{3}$ Dipartimento di Fisica {\em Galileo Galilei}, Universit\`{a} di
Padova, via Marzolo 8, I--35131 Padova, Italy\\
$^{4}$ Dipartimento di Astronomia, Universit\`{a} di Padova, vicolo
dell'Osservatorio 5, I--35122 Padova, Italy\\}

\begin{document}

\maketitle

\begin{abstract}
We consider a weighted biasing scheme for galaxy clustering. This
differs from previous treatments in the fact that the biased
density field coincides with the background mass--density whenever
the latter exceeds a given threshold value. There is some
physical motivation for this scheme and it is in better accord
with intuitive ideas than models based on the Kaiser
(1984) analysis of the clustering of rich clusters.
We explain how different
classes of object could be biased in different ways with
respect to the underlying density distribution but still have
$b=1$. We also show that if one applies our scheme consistently
a weak dependence of $b$ upon density can be implied. This could also
be the reason why the correlation function of galaxies in groups
does not differ substantially from the correlation function of all
galaxies.
\end{abstract}


\section{Introduction}

The assumption that bright galaxies are {\em biased} tracers of
the mass distribution has featured strongly in theories
of galaxy and structure formation in recent years. The
simplest version of the Cold Dark Matter model (CDM) was found
to be unable to explain the clustering properties of galaxies
if the number density of galaxies and the density of
gravitating material were simply proportional to each other
\cite{defw}. The observation by Kaiser (1984) that
the strong clustering of rich clusters relative to galaxies
could be a simple consequence of the fact that clusters
form only in regions where the mass density is particularly
large, led to the adoption of a simple phenomenological model
for {\em biased} galaxy formation wherein galaxies themselves
form only at high peaks of the density field. Because of the
properties of Gaussian statistics -- the simplest and best--motivated
form for the distribution of primordial density fluctuations --
high peaks have different statistical properties
to `typical' points \cite{ph85,bbks} so galaxies
are biased tracers of the mass density field.

The Kaiser model offers a plausible phenomenological explanation
for the clustering of clusters because clusters are defined
to be high density regions in the matter distribution. The
applicability of this model to galaxies themselves is less
clear, however, because in order for it to work one must
suppress galaxy formation from peaks of lower density.
Various authors \cite{r85,s85,ds86,dr87}
have offered some suggestions as to how
a {\em local~} (or {\em natural~})
bias might arise. For example,
the initial density
of a collapsing proto--object might somehow affect the
star formation rate so that low--density peaks form
insufficient stars to produce an identifiable galaxy.
Indeed, given the complexity of the galaxy formation
process it would be surprising if there were not
some kind of non--linear dependence of the space density
of galaxies upon the density of the proto--galaxies from
which they formed \cite{c93}. On the other hand, Peacock (1990) has
found no evidence for such a local bias in the
properties of elliptical galaxies. There is also no strong
trend of clustering amplitude with galaxy luminosity
\cite{wtd88,h88,e89,vab89}, which one would expect if the bias is
controlled by star formation efficiency.  Furthermore, the
recent discovery
by the COBE team of fluctuations in the
sky temperature of the cosmic
microwave background (Smoot et al. 1992) suggests that
galaxy formation in the CDM model should be unbiased
(or even anti--biased so that galaxies form preferentially
in low--density regions).  There is also the possibility that
any bias is not even local: feedback from the first
generation of objects may suppress or enhance galaxy formation
on a large scale around them. Indeed such a non--local bias
seems to be necessary to reconcile standard CDM with
observations of very large scale galaxy clustering \cite{bw91,bcfw}.

So how can one square the observational evidence against biasing
with the theoretical motivation for it? The answer may
well lie in the naivety of the theoretical modelling.
In this paper we shall look at the question of
modelling locally--biased galaxy formation. We shall argue that
the Kaiser model is probably not appropriate for galaxy
clustering and suggest a simple alternative which is in
better accord with intuition. In our scheme -- the
weighted biasing scheme -- the biased density is proportional to the
mass density but
only where the mass density exceeds some threshold value.

The plan of the paper is as follows. In Section 2 we introduce the weighted
biasing scheme, while in Section 3 we present the results for the two--point
correlation of the biased density field (technical details are given
in Appendix A). Section 4 contains a general discussion.

\section{A weighted biasing scheme}

Here we construct a mathematical definition of
our weighted biasing scheme. The underlying mass density
fluctuation field will be denoted
$\delta_{R}$, where $R$ is some smoothing scale used to
define the mass scale of a proto--object:
\begin{equation}
\delta_{R}({\bf x}) \equiv [\rho_{R}({\bf x}) -
\langle \rho \rangle]/\langle \rho \rangle\;.
\end{equation}
The $\langle \; \rangle$ brackets denote averages over a
probability distribution, and we shall assume throughout
that these are equivalent to averages over a large volume
of space i.e. that the field $\rho_{R}({\bf x})$ is
{\em ergodic} (Adler 1981).

In a local bias model, the {\em biased density field} (BDF)
is a local function of $\rho_{R}({\bf x})$. That is to say the probability
of finding a galaxy at a point ${\bf x}$ is a function only of
$\rho_{R}({\bf x})$ or, equivalently, of $\delta_{R}({\bf x})$:
we take $ dP({\bf x}) = n_{\circ} f(\rho_{R})\,dV/\langle f\rangle$
where $n_{\circ}$
is the mean number density of galaxies. General constraints on this
type of model are derived by Coles (1993). Because we
shall be assuming the existence of some threshold
$\nu$, we define the BDF to be $f(\rho_R)=\rho_{\nu, R}({\bf x})$.
If $f(\rho_R)\propto\rho_R$ then galaxies trace the mass (in a
statistical sense) and are therefore `unbiased'.
In the simplest model of biasing, suggested by Kaiser (1984),
the BDF
is defined as constant above a threshold and zero below it, i.e.
\begin{equation}
\rho_{\nu, R}({\bf x}) \equiv
\langle \rho \rangle \,\Theta\!\left( \delta_{R}({\bf x}) - \nu
\sigma_{R} \right)\;;
\end{equation}
the threshold is $\nu \sigma_{R}$ and $\sigma_{R}$ is the
{\em rms} value of $\delta_{R}$. The function
$\Theta(x)$ is the Heaviside step function.
We shall henceforth refer to this model as the Kaiser model.

Notice that the Kaiser model possesses the property that
for very low (i.e. negative)
thresholds $\nu\sigma_R$ galaxy formation occurs with
constant probability throughout space. This limit of the Kaiser
model therefore produces an {\em unclustered} distribution of
objects rather than an {\em unbiased} one.  Similarly, there
is a constant probability of galaxy formation above the
threshold. This implies a one--to--one correspondence
between high peaks and galaxies, which may
be reasonable if the threshold is
very high (so that the galaxies are strongly biased) but
a weak bias (which is what the observations
imply) should merely modulate the spatial distribution
of galaxies relative to the mass distribution. One also
has the problem that merging and disruption
of proto--objects in the non--linear regime
is expected to destroy the correspondence between
high peaks in the initial density field and subsequent
bound structures (Coles et al. 1993). It is difficult therefore
to justify the Kaiser model as a description of biased galaxy
formation.

Our proposed
weighted biasing scheme differs from the Kaiser model in that,
above the threshold (which we take to be the same:
$\nu \sigma_{R}$), the BDF coincides with the underlying density
field:
\begin{equation}
\rho_{\nu, R}({\bf x}) \equiv
\rho_{R}({\bf x}) \,\Theta\!\left( \delta_{R}({\bf x}) - \nu
\sigma_{R} \right)\;.
\end{equation}
In the limit
of small $\nu\sigma_R$, galaxies trace the mass exactly and are
therefore unbiased. For larger $\nu\sigma_R$
the weighted BDF traces the mass, but only in regions where
the density exceeds some critical value. We
do not therefore require a one--to--one correspondence between
peaks and galaxies. Not only does this model agree better with
our intuition in this limit, but it also agrees qualitatively
with the results obtained from hydrodynamical simulations
by Cen \& Ostriker (1992). Indeed, there is good evidence
for the existence of similar density--thresholding behaviour of
star formation in the disks of spiral galaxies
where the local gas density seems to provide the trigger
\cite{g87}.

Note that $\rho_{\nu, R}({\bf x})$
actually represents the true mass--density inside a smoothed sphere of
filtering radius $R$ centered on ${\bf x}$ for those mass fluctuations which
exceed the threshold. We define the ratio
\begin{equation}
\mu_{R}(\nu) \equiv \frac{\langle \rho_{\nu,R} \rangle}{\langle \rho
\rangle}\;,
\end{equation}
which is the total mass fraction in {\it excursion regions}
(i.e. those regions where the mass fluctuation $\delta_{R}$ exceeds the
threshold $\nu \sigma_{R}$). Using the ergodic property, we can
take the expectations to correspond to spatial averages over
a large (formally infinite) volume $V$. It is easy thus to see
that
\begin{equation}
\mu_{R}(\nu) = \frac{ \int_{V} d^{3} {\bf x}\, \rho_{R}({\bf x}) \,\Theta
\!\left( \delta_{R}({\bf x}) - \nu
\sigma_{R} \right)}{\int_V d^{3}{\bf x}\, \rho_{R}({\bf x})} =
\frac{\int_{V_{R}(\nu)}
d^{3}{\bf x}\, \rho_{R}({\bf x})}{\int_V d^{3}{\bf x}\, \rho_{R}({\bf x})}
\equiv \frac{M_{R}(\nu)}{M_{_{TOT}}}\;,
\end{equation}
where $V_{R}(\nu)$ is the total volume of the excursion regions.

In the Kaiser and weighted bias models -- and indeed in
any local biasing scheme involving
a threshold -- one could define the total number of objects
$N_{obj}(>\nu)$
as the total mass above the threshold $M(>\nu)$ divided
by the mass of a single object $M_{obj}$. Thus
\begin{equation}
N_{obj}(>\nu) = \frac{M(>\nu)}{M_{obj}}
= \frac{M(>\nu)}{M_{TOT}}\left(\frac{M_{TOT}}{M_{obj}}\right)\;,
\end{equation}
so that $N_{obj}(>\nu) = \mu_{R}(\nu) M_{TOT}/M_{obj}$. One is implicitly
assuming that the objects form only out of the mass within the
excursion set. If one allows for accretion, however, one could get
a different expression. The quantity $\mu_{R}(\nu)$ is interesting in any
case, however, because it illustrates one of the dangers of adopting
the assumption of Gaussian statistics for the mass fluctuation field
\cite{k84,ph85,bbks,c93}. In such a case
\begin{equation}
\mu_{R}(\nu) = \frac{1}{2}\,{\rm erfc}\left(\frac{\nu}{\sqrt{2}} \right)
+ \frac{\sigma_{R}}{\sqrt{2\pi}}\, {\rm e}^{-\nu^2/2} \equiv \left[\Phi(\nu)+
\nu\sigma_{R} \right] \frac{{\rm e}^{-\nu^2/2}}{\sqrt{2\pi} \,\nu}\;,
\end{equation}
where ${\rm erfc} \,x \equiv (2/\sqrt{\pi}) \int_x^\infty d y
{\rm e}^{-y^2}$ is the
complementary error function and the auxiliary function
$\Phi(\nu)$ has a convenient asymptotic expansion:
\begin{equation}
\Phi(\nu) \equiv \sqrt{\frac{\pi}{2}}\, \nu\, {\rm e}^{\nu^2/2} \,
{\rm erfc}\left(\frac{\nu}{\sqrt{2}}\right)
\simeq 1 +
\sum_{n=1}^\infty (-1)^n \,\frac{(2n-1)!!}{\nu^{2n}}\;.
\end{equation}

We plot the function $\mu_{R}(\nu)$ for different values of
$\sigma_{R}$ in Figure 1. As expected,
the mass fraction $\mu_{R}(\nu)$ tends to unity as $\nu \to -\infty$,
since in this case the BDF reduces to the mass density field.
Note, however, that the maximum value $\mu_{R}(\nu_m)>1$,
is reached at  $\nu_m = - 1/\sigma_{R}$. The reason for this
apparent paradox is that the underlying distribution
permits the existence of regions with negative mass (i.e. $\rho_{R}<0$).
This point can be made more clear by writing eq (5) in terms of
integrals over the probability distribution rather than spatial
averages. The numerator in (5) becomes
$\int_{\langle \rho_R \rangle(1+\nu\sigma_{R})}^\infty
d\rho_{R}\, \rho_{R} \,p(\rho_{R})$, while the denominator
can be split into two parts:
$\int_{-\infty}^{\langle \rho_R \rangle
(1+\nu\sigma_{R})} d\rho_{R} \,\rho_{R}\, p(\rho_{R}) +
\int_{\langle \rho_R \rangle(1+\nu\sigma_{R})}^\infty d\rho_{R}
\,\rho_{R}\, p(\rho_{R})$.
The integral $\int_{-\infty}^{\langle \rho_R \rangle
(1+\nu\sigma_{R})} d\rho_{R} \,\rho_{R}\, p(\rho_{R})$ in the
denominator always contains a negative contribution coming
from $\rho_{R}<0$ events, which is indeed maximised (in absolute
value) when $\nu=-\,1/\sigma_{R}$. The appropriate quantity in the
Kaiser model (which is the volume fraction of
the excursion regions) does not have any factors of $\rho_R$ in the
integrals and is therefore immune to this effect.

\begin{figure}
\vspace{10cm}
\caption{Mass fraction $\mu_{R}$ as a function of $\nu$
for the weighted bias model assuming Gaussian (left) and
lognormal (right) statistics for various values of $\sigma_R$.}
\end{figure}

This illustrates an important problem with the
modelling of galaxy clustering using Gaussian statistics:
one must be very careful to ensure that
unphysical events with $\rho_R<0$ do not contribute to the
statistics of the biased field. In the weighted biasing scheme,
we can consistently apply our model with Gaussian statistics
provided we restrict ourselves either to $\sigma_{R} \ll 1$,
which reduces the probability of unphysical events,
or to $\nu\gg 1/\sigma_{R}$, so that negative mass events have
little effect on statistical properties such as $\mu_{R}(\nu)$.
Different local biasing functions would lead to different
constraints to be satisfied if the model is to be physically
reasonable.

Alternatively, to get a fully self--consistent model
we could use a distribution which
only allows $\rho_{R}>0$. We have therefore also
computed $\mu_{R}(\nu)$ for an underlying {\it lognormal}
distribution of density fluctuations (Coles \& Jones 1991),
which is a simple phenomenological model for the
non--linear density field:
\begin{equation}
p(\delta_{R})= \frac { (1+\delta_{R})^{-1} } { \sqrt {2\pi\, \log
(1+
\sigma_{R}^2)} }\, \exp\left( - \frac { \log^{2}\Bigl[(1+
\delta_{R})\sqrt {1+\sigma_{R}^2}\,\Bigr] }
{ 2\, \log(1+\sigma_{R}^2) } \right) \;.
\end{equation}
Now we find
\begin{equation}
\mu_{R}(\nu) = \frac {1} {2}\,{\rm erfc}\left( \frac { \log \Bigl[(1+\
\nu\sigma_{R})/\sqrt {1+\sigma_{R}^2}\,\Bigr] }
{ \sqrt { 2\, \log(1+\sigma_{R}^2)} }\right)\; .
\end{equation}
As expected, this takes its maximum value $\mu_{R}(\nu_m)=1$ at
$\nu_m=-\,1/\sigma_{R}$, corresponding to the unbiased case and there
are no problems with $\mu_{R}$ exceeding unity. The behaviour
of $\mu_{R}$ as a function of $\nu$ is illustrated also in Figure 1.

\section{The two--point correlation function of the biased density
field}

We shall now turn our attention to the BDF two--point correlation function,
which is defined as
\begin{equation}
\xi_{\nu,R}(r) = \langle \delta_{\nu,R}({\bf x}_1)
\,\delta_{\nu,R}({\bf x}_2)\rangle\;,
\end{equation}
where $r\equiv \vert {\bf x}_1 -{\bf x}_2 \vert$ and
$\delta_{\nu,R}$ is the BDF fluctuation:
\begin{equation}
\delta_{\nu,R}({\bf x}) =
\mu_{R}(\nu)^{-1} \left[1 + \delta_{R}({\bf x}) \right] \,
\Theta\!\left( \delta_{R}({\bf x}) - \nu \sigma_{R} \right) - 1.
\end{equation}
One thus obtains
\begin{equation}
\mu_{R}^2(\nu)\left[ 1 + \xi_{\nu, R}(r) \right] =  \nonumber
\langle \left[1 + \delta_1
 + \delta_2
+ \delta_1 \delta_2 \right]
\Theta\!\left( \delta_1 - \nu \sigma_{R} \right)
\Theta\!\left( \delta_2 - \nu \sigma_{R} \right) \rangle \;,
\end{equation}
where $\delta_i=\delta_R({\bf x}_i)$. The expression (13)
can be written
\begin{equation}
\mu_{R}^2(\nu)\left[ 1 + \xi_{\nu, R}(r) \right]
=\int_{\nu\sigma_{R}}^{\infty}\int_{\nu\sigma_R}^{\infty}
d\delta_1 d\delta_2 \,
[1+\delta_1 +\delta_2 +\delta_1\delta_2 ]
\,p\, (\delta_1,\delta_2; \xi_R)\;.
\end{equation}
For a Gaussian field, $p\,(\delta_1,\delta_2\,; \xi_R)$
is a bivariate
Gaussian distribution with
covariance $\xi_R$ (and variance $\sigma_{R}^{2}$).
One can further express the integral
(14) in terms of integrals over derivatives
of a bivariate Gaussian distribution. The details are given
in Appendix A. For this phenomenological discussion it
is sufficient to discuss the {\em linear} biasing factor:
\begin{equation}
\xi_{\nu,R}(r) \simeq b_{R}^2(\nu)\, \xi_{R}(r),
\end{equation}
obtained by Taylor expanding (14) up to first order in
$\omega_R(r)=\xi_{R}(r)/\sigma_{R}^{2}$. The result is
\begin{equation}
b_{R}(\nu) = \frac{\sigma_{R} \Phi(\nu) +
\nu (1 + \nu \sigma_{R})}{\sigma_{R} \Phi(\nu) + \nu \sigma_{R}^2 }.
\end{equation}
The behaviour of this function is displayed in Figure 2.
Of course, this parameter only describes the bias on large
scales when $\xi_{R}$ is small. In general the bias
$(\xi_{\nu, R}(r)/\xi_{R}(r))^{1/2}$ will be a
non--increasing function of $r$ for any local bias model
(Coles 1993). Galaxy clustering on scales where $\xi$ is not small
is surely dominated by non--linear dynamical evolution
which we have no hope of modelling in a simple way, so
we shall not place any emphasis on the behaviour for large
$\omega_R$. The behaviour of the function $b_{R}(\nu)$ in the weighted
biasing scheme contrasts with the Kaiser (1984) expression,
\begin{equation}
b_{R}'(\nu) = \frac{\nu}{\sigma_{R} \Phi(\nu)}.
\end{equation}
Note that our biasing factor reduces to unity in
the unbiased Gaussian case, $b_{R}(-\infty) = 1$, unlike the Kaiser
expression; like the Kaiser model it gives $b_{R}(\nu)
\simeq \nu/\sigma_{R}$ in the  high--$\nu$ limit.

\begin{figure}
\vspace{20cm}

\caption{Linear biasing factors as a function of $\nu$
in the weighted bias model (top) and
Kaiser model (bottom) for underlying Gaussian statistics
(left) and lognormal statistics (right).
The lines refer to the same values of $\sigma_R$ as those in
Figure 1.}
\end{figure}

Notice, however, that there is another way
to obtain $b=1$ in this version
of the weighted biasing scheme, by choosing
$\nu=\sigma_R-1/\sigma_R$. For any $\sigma_R$, it therefore seems
that there are two values of $\nu$ which can
lead to an `unbiased' model.
However, the minimum of $b_R$ corresponds exactly to the regime
where $\mu_{R}>1$ so one might suspect it to be related to
the presence of negative mass events. To demonstrate this,
we have also computed the linear bias factor for the lognormal
distribution:
\begin{equation}
b_{_{LN}}(\nu) = 1 +  \frac{\tilde{\nu}}{\tilde{\sigma}_R\,
\Phi(\tilde{\nu})},
\end{equation}
where
\begin{equation}
\tilde{\nu}=\frac { \log \Bigl[(1+\nu\sigma_R)/\sqrt{1+\sigma_R^2}\Bigr]}
{\sqrt{2\log (1+\sigma_{R}^{2})}},
\end{equation}
and
\begin{equation}
\tilde{\sigma}_R = \sqrt{\log(1+\sigma_{R}^{2})}.
\end{equation}
The behaviour of this function is also shown in Figure 2. Notice
that one reproduces the Gaussian results in the limit
$\sigma_{R}^{2}
\rightarrow 0$, but the bias is a much flatter function of $\nu$
at large $\sigma_R$ than for the Gaussian case and is
monotonically increasing. This shows that it is indeed the case
that negative mass events distort the calculations for an
underlying Gaussian field. To use our phenomenological model
in situations where $\sigma_R$ is not vanishingly small, we
must therefore incorporate an underlying distribution in which
$\rho_R>0$ everywhere, such as the lognormal (9).
Intruigingly, we find that in the Kaiser model for
lognormal fluctuations we have simply:
\begin{equation}
b_{_{LN}}'(\tilde{\nu}) = b_{_{LN}}(\tilde{\nu}) - 1.
\end{equation}
This again shows that $b\rightarrow 0$ in the Kaiser model
in the limit of small threshold.

\section{Discussion \& Conclusions}

Our weighted bias model provides a simple phenomenological
description of biased galaxy formation, in which galaxies
trace the mass where the local density exceeds some
threshold value. It is of course a simplistic model,
but is reasonably plausible and probably an improvement
upon the Kaiser model in that its behaviour in the limit
of small threshold is to give an unbiased distribution
rather than an unclustered one.

The first important point to emerge from this study is that
the statement that $b=1$ is not equivalent to the statement
that galaxies trace the mass. The $b=1$ version of the
Kaiser has a very different relationship of galaxies to underlying
mass than does the $b=1$ weighted bias model. In the latter
model, galaxies do trace the mass (in a well--defined statistical
sense). In the former they do not. In other local bias models
the relationship is different still. The evidence from
COBE that $b\simeq 1$ in a CDM model does not therefore mean that
a local bias of some form is excluded. Moreover, different
populations of objects could have very different biasing
functions but still have the same value of $b$. Perhaps
this is the reason why spirals and ellipticals, though
known to have different relative abundances in regions
of different density \cite{d80}, seem to have similar levels of bias.

Even if one supposes that galaxy formation is unbiased, our
model finds a possible application in the behaviour of
{\em groups} of galaxies. Ramella et al. (1990) have
found, perhaps surprisingly, that the correlation function
of galaxies in groups is similar to that of all galaxies
(though see also Jing \& Zhang 1988; Maia \& da Costa 1990).
If one takes groups to be defined as regions where
the local density exceeds some threshold, then our
weighted biasing prescription should apply to the
correlation function of galaxies within the groups,
assuming galaxies trace the mass there. Our model
therefore provides a plausible explanation for why
the two correlation functions are similar.

Our weighted bias model cannot produce an anti--bias. In
order to do this one has to have a biasing function
which is a decreasing function of $\rho$; models that
achieve this are given by Coles (1993). (The Kaiser model
produces an anti--bias by virtue of the unnatural assumption
that, in the limit of small thresholds, objects are
seeded everywhere with constant probability.) Obviously,
in view of the argument of the previous paragraph, there
is no contradiction {\em in principle} with,
for example, spirals being suppressed in high--density
regions and still having $b=1$.

We have also demonstrated explicitly that one must be very
careful in using arguments based on Gaussian statistics
to describe clustering in the regime
where $\sigma_R \simeq 1$, even in a purely phenomenological way.
Even the Kaiser model, which is not directly distorted by
events with $\rho_R<0$ for reasons explained
in Section 2, involves a logical inconsistency in this
regime. Any bias in galaxy clustering is expected to be
imposed in the non--linear regime and the appropriate
statistical distribution must reflect that fact. If the
bias is imposed at very late times, when the distribution is
highly skewed, then one can get a bias which depends
very slowly on threshold. This could well be the
explanation for why there seems to be little
difference in bias for objects of widely--varying (presumed)
initial density.

\section* {Acknowledgments}
We all thank Francesco Lucchin for discussions;
Paolo Catelan further thanks Manolis Plionis and Enzo Branchini.
Peter Coles receives an SERC Advanced Fellowship; he is also grateful
to the Dipartimento di Astronomia, Universit\`{a} di Padova
for their hospitality during a visit when this paper was
written and to the Consiglio Nazionale delle Ricerche
for financial support. This work was partially supported
by Italian MURST.

\appendix
\section{Derivation of Correlation Functions}

In this Appendix we  derive an explicit form
for the BDF two--point correlation
function $\xi_{\nu,R}(r)$, as
defined in eq.(14) of the text, for an underlying
Gaussian mass density field. In this case one can define
$\alpha\equiv\delta_1/\sigma_R$, $\beta\equiv\delta_2/\sigma_R$; we then need
$p(\alpha,\beta; \omega_R)$, which is a bivariate Gaussian:
\begin{equation}
p(\alpha,\beta; \omega_R)= \frac{1}{2\pi\sqrt{1-\omega_R^2}}
\exp \left[-\frac{\,\alpha^2+\beta^2 -2\,\omega_R\,\alpha\,\beta\,}
{2\,(1-\omega_R^2)} \right].
\end{equation}
It is straightforward thus to show that
\begin{eqnarray}
I(\nu) \equiv  \mu_R^2(\nu)\Bigl[1+\xi_{\nu,R}(r)\Bigr] & = &
\int_\nu^\infty \int_\nu^\infty \! d\alpha \, d\beta
\left[ (1+\sigma_R^2\,\omega_R)-2\,\sigma_R(1+\omega_R)\,
\frac {\partial}{\partial\alpha}+\right.\nonumber\\
 & & 2\,\sigma_R^2\,\omega_R \, \frac {\partial^2}{\partial\alpha^2} +
\left.  \sigma_R^2(1+\omega_R^2)\, \frac {\partial^2}{\partial\alpha\,
\partial\beta}\right]\, p(\alpha,\beta;\omega_R).
\end{eqnarray}
Some of the integrals in (A2) can be performed exactly and
we get the expression
\begin{eqnarray}
I(\nu) & = &
\Bigl[1+\omega_R(1+\nu\sigma_R)\Bigr]\frac {\sigma_R\,
{\rm e}^{-\nu^{\,2}/2}}
{\sqrt {2\pi}}\,\,{\rm erfc}\left(\frac {\nu}{\sqrt {2}}\sqrt {\frac
{1-\omega_R}
{1+\omega_R}}\right)+\nonumber\\
& & \frac{1+\sigma_R^2\,\omega_R}
{2\sqrt{2\pi}}
\int_{\nu}^{\infty} \! d\alpha\, {\rm e}^{-\alpha^{\,2}/2}\, {\rm erfc}
\left(\frac {\nu-\alpha\,\omega_R}{\sqrt {2(1-\omega_R^2)}}\right)
+ \frac {\sigma_R^2 \sqrt {1-\omega_R^2}}{2\pi}\,
\,{\rm e}^{\,-\nu^{\,2}/(1+\omega_R)}\;.
\end{eqnarray}
The expression for the standard bias (Kaiser 1984) is easily recovered
from the latter equation, by taking the $\sigma_R\rightarrow 0$ limit at
fixed $\nu$ and $\omega_R\,$:
\begin{equation}
1 + \xi_{\nu{_R}}'(r) = \frac{\sqrt{2/\pi}}{[{\rm erfc}(\nu/\sqrt{2})]^2}
\int_\nu^\infty d \alpha\,
{\rm e}^{-\alpha^2/2}\,{\rm erfc}\left(\frac{\nu-\alpha\,\omega_R}{\sqrt{
2(1 - \omega_R^2)} } \right).
\end{equation}
We also recover the Politzer \& Wise (1984) expression
in this limit:
\begin{equation}
1 + \xi'_{\nu,R} \simeq \exp(\nu^2\omega_R).
\end{equation}

The result (A3) leads to an exact expression for the BDF
variance in the limit of zero lag, $\omega_R\rightarrow 1$:
\begin{equation}
\sigma_{\nu,R}^2=\frac {(1+\sigma_R^2)\Phi(\nu)+\nu\sigma_R(2+\nu\sigma_R)}
{\left[\Phi(\nu)+\nu\sigma_R\right]^2({\rm e}^{\,\nu^{\,2}/2}
\sqrt {2\pi}\,\nu)}-1\;,
\end{equation}
which in the limit $\nu\rightarrow -\,\infty$ reduces to the background mass
variance, $\sigma_R^2$ (again, unlike the Kaiser model).

The linear weighted biasing factor (16), as explained in the text,
can be obtained in the limit of small correlations by
Taylor expanding (A3) up to first
order in $\omega_R$. This involves approximations such as
\begin{equation}
\exp\left(\,-\frac{\nu^{2}}{(1+\omega_R)}\right)\simeq
(1+\nu^2\omega_R)\,{\rm e}^{-\nu^{\,2}}
\end{equation}
and
\begin{equation}
{\rm erfc}\left(\frac {\nu}{\sqrt {2}}\sqrt {\frac {1-\omega_R}
{1+\omega_R}}\right)\simeq {\rm erfc}\left(\frac{\nu}{\sqrt {2}}\right)
+\sqrt {\frac {2}{\pi}} \,\nu\,\omega_R\,{\rm e}^{-\nu^{\,2}/2}.
\end{equation}
We thus obtain equation (16).

\end{document}